# Author's Accepted Manuscript

This is the version of the article accepted for publication in SUMMA 2024 after peer review. The final, published version is available at IEEE Xplore.

To cite this article, please use the following official publication details:

A. Lazarev and D. Sedov, "Utilizing Modern Large Language Models (LLM) for Financial Trend Analysis and Digest Creation," 2024 6th International Conference on Control Systems, Mathematical Modeling, Automation and Energy Efficiency (SUMMA), Lipetsk, Russian Federation, 2024, pp. 317-321, doi: 10.1109/SUMMA64428.2024.10803746

DOI: https://doi.org/10.1109/SUMMA64428.2024.10803746

NLP models and algorithms applicable to the analysis of financial texts.

Therefore, this research aims to demonstrate the capabilities of NLP in the context of financial trend analysis, which can find widespread use among both professional analysts and individual investors seeking more informed and timely decision-making.

## II. JUSTIFICATION FOR ANALYZING TRENDS AND PUBLICATIONS

Today the volume of publicly available information is growing very rapidly."IDC estimates that the amount of the global datasphere subject to data analysis will grow by a factor of 50 to 5.2ZB in 2025; the amount of analyzed data that is "touched" by cognitive systems will grow by a factor of 100 to 1.4ZB in 2025" [1]. This includes scientific articles, reports, news articles, analytical reviews, and social media posts. To make justified decisions, especially in fields like finance, business, and technology, it's crucial to identify and analyze current trends promptly. However, traditional methods of analyzing publications, which involve manual review and processing, are becoming increasingly inefficient. Here are several key aspects that highlight the need for automated analysis of trends and publications:

### A. Volume and velocity of information

Data Volume: millions of articles, reports, and news items are published daily. Manually tracking and analyzing such a vast amount of information is simply infeasible.

Speed of updates: trends and market conditions change rapidly, and identifying changes promptly can provide a significant competitive advantage, but without automated analysis, information may become outdated before it is even reviewed.

### B. Data complexity and the need for structuring information

Variety of sources: Information comes from a multitude of sources, including scientific journals, blogs, news websites, and social media platforms. Each source may present data in different formats, making integration and analysis difficult.

Unstructured data: A significant portion of valuable information exists in textual form (unstructured data), requiring the application of sophisticated Natural Language Processing (NLP) techniques for analysis and structuring.

*Abstract—*

**The exponential growth of information presents a significant challenge for researchers and professionals seeking to remain at the forefront of their fields and this paper introduces an innovative framework for automatically generating insightful financial digests using the power of Large Language Models (LLMs), specifically Google's Gemini Pro.**

**By leveraging a combination of data extraction from OpenAlex, strategic prompt engineering, and LLM-driven analysis, we demonstrate the automated example of creating a comprehensive digests that generalize key findings, identify emerging trends. This approach addresses the limitations of traditional analysis methods, enabling the efficient processing of vast amounts of unstructured data and the delivery of actionable insights in an easily digestible format.**

**This paper describes how LLMs work in simple words and how we can use their power to help researchers and scholars save their time and stay informed about current trends. Our study includes step-by-step process, from data acquisition and JSON construction to interaction with Gemini and the automated generation of PDF reports, including a link to the project's GitHub repository for broader accessibility and further development.**

## I. INTRODUCTION

In recent years, Natural Language Processing (NLP) technologies have made significant advancements, opening new possibilities for analyzing large volumes of textual data. One of the most promising areas of NLP application is the analysis of current trends and publications, especially in dynamic and information-rich fields such as finance and related areas. Financial markets are characterized by a vast and complex information flow, mixing everything from breaking news and analytical reviews to sophisticated scientific articles. While traditional market data plays a crucial role, academic research adds a layer of depth and these articles, authored by economists, mathematicians, and computer scientists, explore intricate market dynamics, new algorithms, and emerging trends often invisible to the naked eye.

This work aims to explore the application of NLP methods for analyzing current trends and publications, on an example of creation of a financial digest generated entirely by large language model (LLM) Gemini Pro. We also review main

## C. Supporting data-driven decision making

Data-driven decisions: analyzing trends and publications with NLP facilitates making decisions based on objective data and up-to-date information rather than intuition or outdated analysis methods.

Implementing an evidence based approach: In various fields, such as finance, medicine, and science, an evidence-based approach requires constant monitoring and evaluation of new data to ensure the relevance of decisions made.

## III. THE ROLE OF NLP MODELS IN AUTOMATIC KEY INFORMATION EXTRACTION

With the advancement of Natural Language Processing (NLP) technologies, automatic extraction of key information from textual data has become a critical area in artificial intelligence. These models enable the processing of large volumes of data, automatically extracting essential insights that can be used for analysis, decision-making, or generating new knowledge. This chapter explores the main approaches and techniques used by NLP models for automatic key information extraction, providing an overview of scientific research and its findings.

### A. Named Entity Recognition (NER)

One of the primary tasks in information extraction is Named Entity Recognition (NER), which involves identifying and classifying mentions of people, organizations, locations, dates, and other key entities in text. NER models are trained on annotated data, enabling them to automatically identify and classify entities in the text. This process is crucial in applications such as automatic text annotation, information retrieval, and data organization.

Examples in Table I. illustrate how NER can be applied across various domains to extract structured information from unstructured text, making it easier to analyze and understand the content.

TABLE I. NER EXAMPLES

| Example Sentence | Named Entity | Term |
|---|---|---|
| Senator Tim Jones gave a speech at the White House, US on January 3, 2018, regarding Microsoft's new policies. | Person | Tim Jones |
| | Location | White House, US |
| | Organization | Microsoft |
| | Date | January 3, 2018 |
| Elon Musk commented on Twitter that the research conducted by MIT demonstrated the effectiveness of CRISPR-Cas9 in gene editing. | Person | Elon Musk |
| | Organization | MIT |
| | Organization | Twitter |
| | Technology | CRISPR-Cas9 |

In research study "A survey of Named Entity Recognition and Classification" David Nadeau and Satoshi Sekine write that "NERC will have a profound impact in our society. Early commercial initiatives are already modifying the way we use yellow pages by providing local search engines (search your neighborhood for organizations, product and services, people, etc.). NERC systems also enable monitoring trends in the huge space of textual media produced every day by organizations, governments and individuals. It is also at the basis of a major advance in biology and genetics, enabling researchers to search the abundant literature for interactions between named genes and cells." [2]

### B. Automatic Text Summarization

Text summarization, a key application of Natural Language Processing (NLP), aims to break lengthy documents into brief summaries without sacrificing essential information. This process, utilized by news aggregators, legal researchers, and others dealing with large volumes of text has two primary techniques: extractive summarization, which identifies and extracts key sentences from the original text, and abstractive summarization, which leverages NLP models to paraphrase and generate new sentences that capture the document's central themes. Both approaches strive to significantly reduce reading time while preserving the core message and meaning of the source material. [3, 4]

*1) Extractive Summarization:* Extractive summarization works by identifying and selecting the most important sentences or phrases directly from the source text. The selected segments are then compiled into a summary. This method relies heavily on statistical and linguistic features, such as word frequency, sentence position, and the presence of specific keywords. Extractive summarization is often preferred due to its simplicity and the fact that it typically results in grammatically correct summaries because it uses actual sentences from the original text.

*2) Abstractive Summarization:* Abstractive summarization, on the other hand, involves generating new sentences that capture the main ideas of the source text, rather than just extracting segments. This method mimics how a human might summarize a document, by understanding the context and generating new text that conveys the same meaning but with fewer words. Abstractive summarization is more challenging due to the need for advanced understanding and generation capabilities, and it often involves techniques from deep learning, such as sequence-to-sequence models and transformers.

Here in Table II below is an example of how summarization works and differences the models have between each other.

TABLE II. ABSTRACTIVE SUMMARIZATION EXAMPLES

| Sentence | Summarization | Result | Explanation |
|---|---|---|---|
| A powerful earthquake struck central California on Tuesday, causing widespread damage to buildings and infrastructure. The magnitude 6.2 quake hit at 2:15 p.m. local time, according to the U.S. Geological Survey. Its epicenter was located about 10 miles north of the town of Hollister, which suffered significant damage. Rescue workers are on the scene assessing the damage and searching for potential survivors. The governor has declared a state of emergency for the affected areas. | Abstractive | A strong earthquake hit central California, with the epicenter near Hollister, causing extensive damage and prompting a state of emergency declaration. Rescue efforts are underway. | Abstractive summaries rephrase information, often using different words and sentence structures. Abstractive models aim for a more fluent and human readable summary and this method is more complex, requiring advanced NLP models. It can be exposed to factual errors if the model misinterprets the source text. |
| | Extractive | A magnitude 6.2 earthquake struck central California on Tuesday, causing widespread damage. Its epicenter was located about 10 miles north of Hollister, which suffered significant damage. The governor has declared a state of emergency. | In extractive summarization, the most important segments are pulled directly from the text without rephrasing. These model results sometimes sound choppy or lack smooth transitions. This method is generally easier to implement than abstractive summarization and extractive model works best when key information is concentrated in specific sentences. |

## C. Relation Extraction

Relation extraction is part of NLP, which focuses on identifying and categorizing semantic relationships between entities mentioned in the text. This technique enables NLP models not only to recognize individual entities but also to understand how they are interconnected, providing deeper insights into the data. For example, in medical texts, this could be the relationship between a drug and a disease, while in legal documents, it could be the relationship between contract parties. [5]

*1) Types of Relationships in Relation Extraction.*

Relation Extraction typically involves the identification of various types of relationships, including but not limited to:

*a) Binary Relationships*

In Table III there is a type of relations which involve two entities.

TABLE III. ENTITIES RELATIONS EXAMPLES

| Sentence | Entities | Relation Type | Extracted Relation |
|---|---|---|---|
| Mike and John work for Tesla Inc. | Mike (Person), John (Person), Tesla Inc. (ORG) | Employment | Person → ORG |
| Obama was born in Hawaii. | Obama (Person), Hawaii (Location) | Birthplace | Person → Location |

In this case, the relationship is "work for", and the two entities involved are Mike (Person) and Tesla Inc. (Organization). While two people (Mike and John) are mentioned, each forms an individual binary relationship with Tesla Inc. So, for both Mike and John, the relationship "work for" connects a person with the organization.

*b) N-ary Relationships*

More complex relationships that involve more than two entities, such as "patient-disease-treatment" in medical texts are illustrated in Table IV.

TABLE IV. N-ARY RELATIONS EXAMPLES

| Sentence | Entities | Relation Type | Extracted Relation |
|---|---|---|---|
| Dr. Pakler prescribed antibiotic to treat Diana's bacterial infection | Dr. Pakler (Person), antibiotic (Drug), Diana (Person), bacterial infection (Disease) | Medical | Person → Drug → Person → Disease |
| The study found that the combination of aspirin and clopidogrel reduced the risk of stroke in high-risk patients. | aspirin (Drug), clopidogrel (Drug), stroke (Disease), high-risk patients (Group) | Medical | Drug → Disease → Group |

*c) Hierarchical Relationships*

Hierarchical relationships describe parent-child or part-whole relations, they are often seen in classification, ontologies and organizational charts. Most common relations are: "is-a" , "part-of" or "member-of" (group of something) relations. Some examples are shown in Table V.

TABLE V. HIERARCHICAL RELATIONSHIPS EXAMPLES

| Sentence | Entities | Relation Type | Extracted Relation |
|---|---|---|---|
| A rose is a type of flower. | rose (plant), flower (plant category) | classification | is-a |
| Head is a part of the human body. | head (component), body (whole object) | hierarchy | part-of |

*2) Two main approaches to Relation Extraction.*

*a) Supervised Learning Approaches.*

Supervised methods rely on labeled training data to learn the relationships between entities. These models typically use features derived from the text, such as the words between the entities, their positions in the sentence, and dependency parse trees.

*b) Unsupervised and Semi-supervised Approaches.*

These approaches attempt to extract relations without extensive labeled data. Unsupervised methods cluster similar sentences and then infer possible relationships. Semi-supervised methods, like bootstrapping, start with a small amount of labeled data and iteratively expand the training set.

## IV. GEMINI PRO 1.5 MODEL BY GOOGLE

Gemini pro is a LLM (Large Language Model) which was trained on a massive dataset of text and code, incorporating within itself a wide variety of information. This model is capable of generating human-quality text, translating text to different languages, writing fictional or creating content and answering questions.

### A. Key aspects of the model

Gemini is a **Large Language Model** which uses a Transformer-based **Neural Network** and this architecture is good for processing sequential data like language, allowing it to understand context and relationships between words in a sentence or even across paragraphs.

*1) Transformer-based Neural Network:* is a type of artificial intelligence that excels at understanding relationships between words in a sentence, even if they are far apart, it does this using a mechanism called "attention," which allows it to focus on the most relevant parts of the input when generating output. This makes Transformers incredibly powerful for tasks like language translation, text summarization, and building conversational AI models like Gemini.

*2) Attention Mechanism:* this is the heart of a Transformer, it allows the model to focus on specific parts of the input sequence that are most relevant to the task at hand.

Example of Attention Mechanism: In the sentence "The cat sat on the mat," the attention mechanism would understand that "cat" is closely related to "sat" and "mat," even though they are not right next to each other.

## V. FROM THEORY TO ACTION: USING GEMINI AND OPENALEX DATA TO MAKE AI FINANCIAL DIGEST

### A. Why is it important?

In previous sections, we explored the theoretical foundations of neural networks. Now, let's examine how neural networks can empower researchers by keeping them abreast of the latest scientific trends and noteworthy publications.

This digest addresses the challenges of information overload and rapid change in the financial world, it provides concrete and actionable insights of what is going on in this field of study for little effort from the researcher.

By using AI to analyze data and identify patterns before they become mainstream, the digest reveals emerging trends, it helps professionals make informed choices, supporting data-driven decision-making.

### B. Process of creation Financial Digest

There are 3 steps to create such report:

*1) Input Data Processing:* how Gemini model processes the input data sourced from OpenAlex.
*2) Information Extraction and it's generalization.*
*3) Conducting automated report.*

The creation of our AI-powered Financial Digest begins with meticulous data gathering and preparation. Here's a detailed look at **Input Data processing** step:

- Fetch data from OpenAlex's Open Library:

We use the OpenAlex platform, a vast, publicly available database of scholarly literature. Especifically, we extract research article abstracts focused on finance and emerging markets published within a specific timeframe - typically the previous month, with a delay of 2-3 weeks to allow for sufficient data collection and indexing.

- Defining the Scope:

To ensure relevance, we employ targeted queries to pinpoint articles classified under the Field of Study (FOS) categories of "finance". This focused approach ensures we gather information directly relevant to our target audience.

- Storing and Preparing the Data:

The collected abstracts are stored securely and organized for efficient processing in postgres database, we need to make a structured JSON of this data to make it understandable for Gemini.

After we conduct proper JSON for Gemini we can proceed to **Information Extraction and it's generalization** step

- Engaging Gemini with Strategic Prompts:

On this step we send constructed JSON into Gemini large language model, but we don't just provide the raw data – we guide Gemini's analysis through carefully crafted prompts. These prompts act as instructions, directing Gemini to extract specific information and insights for different sections of the Financial Digest.

Example Prompts:

a) "Summarize the key findings in all abstract parts of this article from the provided research articles. Please describe widely and cite most interesting.",

b) "What are the main themes or trends emerging from these articles? Take into account only abstracts (values)",

c) "Can you identify any commonalities or connections between the different research papers?",

d) "What are the major implications for future directions suggested by this research? Describe future possibilities."

- Output integration:

So we send JSON data and Gemini processes the JSON data and prompts, generating human-readable text as output. These outputs full with insights and analysis are stored to form the basis of the different sections of our Financial Digest.

## VI. CONDUCTING PDF DOCUMENT

We **dynamically create a title page** for our Financial Digest using a Python library like ReportLab or PyFPDF. The title page includes: title, period covered (which automatically extracted from your data processing stage) and authors.

**Structuring the Digest:** The outputs generated by Gemini, now forming the sections of our digest, are formatted using headings, subheadings, and bullet points for clarity and readability. We add section breaks and page numbers for better navigation within the PDF document, conduct list of contents and list of sourced articles with their titles and DOIs.

**Incorporating Visualizations (Optional):** If your analysis includes charts or graphs generated from the data, these can be automatically incorporated into the relevant sections of the PDF using libraries like matplotlib or seaborn.

**Saving the PDF:** Finally, the entire document, including the title page, structured content, and optional visualizations, is saved as a PDF file, filename also programmatically created, e.g. "FinancialDigest_YYYY_MM.pdf" to ease identification.

GitHub Repository: The code for this entire process, from data acquisition and JSON construction to Gemini interaction and PDF generation, is available in our public GitHub repository, researchers and developers can explore the code in a GitHub repository [6].

## VII. Conclusion

This research explored the potential of harnessing the power of modern Large Language Models (LLMs), specifically Google's Gemini, to automate the creation of insightful financial digests. In recent years, the volume and complexity of information have grown rapidly [1] and traditional methods of analysis are becoming less efficient while they still require a lot of time.

By automating this process, we created an example of how modern LLM can help generalize data to simplify its consumption and help researchers and scholars do their job. Modern LLMs process vast amounts of unstructured data about Field of Study, identify emerging trends, and synthesize key findings into readily digestible formats, such as PDF.

This work demonstrated a practical framework for realizing this vision. By combining data extraction from open-access repositories like OpenAlex, strategic prompt engineering, and LLM-driven analysis, we can automate the creation of comprehensive financial digests. These digests, enriched with synthesized summaries, trend identification, and potential future implications, offer invaluable support for researchers, investors, and decision-makers navigating the complex world of finance.

The impact of this approach extends beyond the realm of finance. The principles and techniques detailed in this paper can be readily adapted to other domains facing similar challenges of information overload, including scientific research, market analysis, and technology forecasting. The provided GitHub repository serves as an example to researchers and developers to explore.

While LLM technology continues to advance, we can anticipate even more sophisticated capabilities in the future.